\documentclass[conference]{IEEEtran}
\IEEEoverridecommandlockouts

\usepackage{cite}
\usepackage{amsmath,amssymb,amsfonts}
\usepackage{graphicx}
\usepackage{textcomp}
\usepackage{xcolor}
\usepackage{cleveref}
\usepackage{acronym}
\usepackage{algpseudocode}
\usepackage{algorithm}
\usepackage{enumitem}
\usepackage{url}
\usepackage{array}

\def\BibTeX{{\rm B\kern-.05em{\sc i\kern-.025em b}\kern-.08em
    T\kern-.1667em\lower.7ex\hbox{E}\kern-.125emX}}

\def\x{\boldsymbol{x}}\def\p{\boldsymbol{p}}\def\lamlam{\boldsymbol{\lambda}}

\begin{document}

% Define acronyms without displaying them
\newacro{QoS}[QoS]{Quality of Service}
\newacro{BS}[BS]{base station}
\newacro{RB}[RB]{resource block}
\newacro{RBs}[RBs]{resource blocks}
\newacro{OFDMA}[OFDMA]{Orthogonal Frequency Division Multiple Access}
\newacro{CL}[CL]{Capacity Limited}
\newacro{URLLC}[URLLC]{Ultra Reliable Low Latency Communication}
\newacro{TS}[TS]{Time Sensitive}
\newacro{5G}[5G]{5th generation}
\newacro{NS}[NS]{network slicing}
\newacro{RRA}[RRA]{radio resource allocation}

\title{Slice-aware Resource Allocation and Admission Control for Smart Factory Wireless Networks \\ \thanks{R. Ochonu is supported by the 5GSmartFact project, which has received funding from the European Union's H2020 research and innovation program under the Marie Skłodowska-Curie grant ID 956670. J. Vidal's work is funded by the European Commission Horizon Europe SNS JU PREDICT-6G (GA 101095890) Project and the NextGeneration UNICO5G TIMING (TSI-063000-2021-145). Also by the project 6-SENSES grant PID2022-138648OB-I00, funded by MCIN/AEI/10.13039/501100011033, and by FEDER-UE, ERDF-EU \textit{A way of making Europe}, and the grants 22CO1/008248 and 2021 SGR 01033 (AGAUR, Generalitat de Catalunya).}
}

\author{\IEEEauthorblockN{Regina Ochonu, Josep Vidal}
\IEEEauthorblockA{\textit{Dept. of Signal Theory and Communications, Universitat Politècnica de Catalunya, Barcelona, Spain} \\
\{regina.ochonu, josep.vidal\}@upc.edu}}

\maketitle

\begin{abstract}
The \acf{5G} and beyond network offers substantial promise as the ideal wireless technology to replace the existing inflexible wired connections in traditional factories of today. 5G network slicing allows for tailored allocation of resources to different network services, each with unique \acf{QoS} requirements. This paper presents a novel solution for slice-aware radio resource allocation based on a convex optimisation and control framework for applications in smart factory wireless networks. The proposed framework dynamically allocates minimum power and sub-channels to downlink mixed service type industrial users categorised into three slices: \acf{CL}, \acf{URLLC}, and \acf{TS} slices. Given that the \acf{BS} has limited transmission power, we enforce admission control by effectively relaxing the target rate constraints for current connections in the \ac{CL} slice. This rate readjustment occurs whenever power consumption exceeds manageable levels. Simulation results show that our approach minimises power, allocates sub-channels to users, maintains slice isolation, and delivers QoS-specific communications to users in all the slices despite time-varying number of users and changing network conditions.
\end{abstract}


\section{Introduction}
Advanced technologies such as extended reality (XR), robotics and automation, digital twins, and cyber-physical systems, are pivotal in the development of smart factories, and can significantly boost the efficiency of manufacturing processes\cite{b1}\cite{b2}. One of such innovative technologies is network slicing, a key feature of 5G wireless networks that utilise network functions virtualization (NFV) and software-defined networking (SDN) techniques to create logically isolated virtual networks (or 'slices') on top of the same physical infrastructure \cite{b3}. Each slice functions as a distinct, dedicated end-to-end network and can be customised to meet the specific Quality-of-Service (QoS) requirements of different applications in terms of capacity, reliability, latency, and security. For example, one slice could be optimised for industrial IoT devices with low power and bandwidth needs, while another could support high-speed, low-latency applications like virtual reality and autonomous guided vehicle (AGV)\cite{b4}.

In an industrial setting, the complexity of managing shared radio resources is heightened by factors such as mobility of terminals, interference, varying propagation conditions, and diverse traffic types. These elements introduce uncertainty that must be carefully considered when allocating resources to the different slices tailored for various smart factory applications. Ensuring that each slice meets its specific QoS requirements while maintaining strict isolation to prevent issues in one slice from impacting others presents a significant challenge. This is especially true given the demands for sub-millisecond latency and high reliability in the dynamic and interference-rich factory floor. Furthermore, with the constraints posed by limited radio resources \cite{b5} and the potential for high device density in smart factories \cite{b6}, it is essential to deploy effective admission control mechanisms. These mechanisms are vital for managing network congestion and maintaining inter-slice priorities, thereby adding an extra layer of complexity to network resource management.

The problem of allocating radio resources remains a significant practical engineering challenge and thus a continuing research topic in wireless communications\cite{b7}. The effectiveness of a resource allocation algorithm in real-world applications depends on its speed and accuracy in devising an allocation scheme \cite{b6}. Different optimisation-based models have been proposed to solve slice-aware resource allocation problems for specific use cases. In \cite{b8}, a stochastic optimization approach using the Lyapunov drift-plus-penalty method was leveraged to minimize power while ensuring reliable and isolated slice operations. Meanwhile, \cite{b9} explored methods like Big-M formulation and successive convex approximation to allocate power and resource blocks to support different services. The limitations of both frameworks stem from the complexity of their problem formulations, requiring advanced computational techniques for effective implementation. More recently, machine learning-based models have gained significant attention. For example, \cite{b10} and \cite{b11} propose deep reinforcement learning frameworks to dynamically optimize resource allocation for slicing-enabled mixed services. Their approach significantly enhance the autonomous management of resources in the diverse and complex traffic flows of 5G services; however, they are limited by prolonged training and convergence times stemming from the complexity and high-dimensional nature of their learning algorithms. 

In this paper, we propose a slice-aware downlink (DL) resource allocation and admission control strategy based on optimization and control framework for enabling network slicing with effective isolation in a smart factory network environment. The main contributions of this paper are as follows;
\begin{itemize}
    \item Dynamically allocate DL radio resources such as power and subchannels to mixed traffic type users while prioritizing their unique QoS requirements. We model this problem by minimizing transmit power under rates constraints.
    \item Significantly reduce the complexity of the optimisation problem formulation by translating all QoS constraints to target rate constraints. This translation allows for a efficient use of convex optimization algorithms, unlike in \cite{b8}, \cite{b9}.
    \item Ensure slice isolation using a feedback control system for users in the CL slice, close form target rate expressions for users in the URLLC and TS slices, and pre-emption admission control strategy. 
\end{itemize}

The rest of the paper is structured as follows. Section II describes the system model and problem formulation. Section III details the strategy for resource allocation, guaranteed slice isolation and admission control. Simulation results are shown in Section IV. The paper concludes in Section V with summary remarks and potential avenues for future research.

\section{System model}
Consider the DL of an Orthogonal Frequency-Division Multiple Access (OFDMA) cellular network with a single base station (BS) serving a set $\mathcal{N}$ of $N$ users and having a set $\mathcal{K}$ of $K$ available subchannels. Each user belongs to a network slice and has its own, individual rate, reliability, latency, and jitter requirements. At time slot $t$, the total number of users in the system is $N(t)$ and the number of users in a slice $s$ is $N_s(t)$. The maximum arrival rate for each user in slice $s$ follows a Poisson distribution. We assume that the packet lengths for each user follow an exponential distribution. Thus, the packet queues at the BS follow an M/M/1 queuing model.

We classify users into three slice types depending on their specific application and the type of traffic they generate. Many smart factory use cases can be accommodated in these slice categories\cite{b12}\cite{b13}:
\begin{enumerate}
    \item \textit{Capacity limited (CL) slices}. In these slices, only the total capacity of the slice matters, designated for best-effort traffic where the network does not provide any guarantee on the delivery, timing, or sequencing of packets. Smart factory applications such as file sharing, non-critical monitoring data, and web browsing, are classed in the CL slice. Multiple users of this traffic type can be accommodated using available, non-committed resources, as long as their combined data rate remains below the specified slice capacity limit, \(C_{s}\).
    \item \textit{Ultra-reliable low-latency communication (URLLC) slices}. For applications such as augmented reality, real-time control of robots and machinery, AGV, and real-time surveillance/security systems with stringent requirements on latency, jitter and reliability. This is typically asynchronous traffic characterized by a target outage delay given by $\Pr\left(D_i>D_i^{max}\right)<1-\gamma_i$ and short packet lengths, where $\gamma_i$ and $D_i^{max}$ are the target reliability and the maximum allowable delay respectively, for a user $i$ in a URLLC slice.
    \item \textit{Time-sensitive (TS) slices}. These slices cater to applications that demand immediate responses, such as machine vision systems or predictive maintenance. They are designed to handle periodically generated isochronous short packets that must be served upon arrival at the queue, with a target maximum bit error rate (BER). Typically, these packets arrives at the queue at periodic intervals that are multiples of the scheduling period $T_{sp}$, with a fixed packet size $L$.
\end{enumerate}
The network must allocate power and subchannels required for network slicing in a manner that satisfies the QoS requirements of all slice types while ensuring their complete isolation in a dynamic network environment.

The DL transmission rate from the BS to user $i$ in a Gaussian channel is upper bounded by the Shannon formula
\begin{equation} r_i(t)=B\sum_{j=1}^{K}x_{ij}(t)\log_2\left(1+\beta_i\frac{p_{ij}(t)h_{ij}(t)}{x_{ij}(t)\sigma^2}\right),
\label{eq:shannon formula}
\end{equation}
 where $B$ is the subchannel bandwidth and $h_{ij}(t)$ is the Rayleigh fading channel gain of the transmission from BS to user $i$ on subchannel $j$ at time slot $t$. $\sigma^2$ is the noise power. Note that the interference levels across subchannels is captured in $\sigma^2$. $p_{ij}(t)/x_{ij}(t)$ is the DL transmission power on subchannel $j$ to user $i$ at slot $t$. $x_{ij}(t)$ is the subchannel allocation indicator with $x_{ij}(t)=1$ when subchannel $j$ is allocated to user $i$ at time slot $t$, else $x_{ij}(t)=0$. The presence of $x_{ij}(t)$ in the denominator of \eqref{eq:shannon formula} guarantees the joint concavity of $r_i$ with respect to $x_{ij}(t)$ and $p_{ij}(t)$ \cite{b15}. Finally, $\beta_i$ is a factor required to get a tight bound on the uncoded BER for an M-QAM modulation associated to user \textit{i} \cite{b14}. Based on this model, the goal is to minimize the average power while maintaining slice isolation, reliability, latency and rates, a problem formally posed as
\begin{equation}
\begin{aligned}
\min_{p,x} \ & \lim_{t \to \infty} \frac{1}{t} \sum_{\tau=1}^{t} \sum_{i=1}^{N(\tau)} \sum_{j=1}^{K} p_{ij}(\tau), \\
\text{s.t.} \ & \Pr\left(D_i > D_{i}^{\max}\right) < 1 - \gamma_i, \ \forall i \in \mathcal{N}_{s}, \ \forall s \in \mathcal{S}_{D}, \ \forall t,\\
& \lim_{t \to \infty} \frac{1}{t} \sum_{\tau=1}^{t} \sum_{i=1}^{N_s(\tau)} r_i(\tau) = C_s, \ \forall s \in \mathcal{S}_{C}, \\
&r_i(t) = L/T_{sp}, \ \forall i \in \mathcal{N}_s, \ \forall s \in \mathcal{S}_{T} \\
& p_{ij}(t) \geq 0, \ x_{ij}(t) \in \{0,1\}, \ \forall i \in \mathcal{N}, \ \forall j \in \mathcal{K}, \ \forall t, \\
& \sum_{i=1}^{N} x_{ij}(t) \leq 1, \ \forall j \in \mathcal{K}, \ \forall t,
\end{aligned}
\label{eq:min_problem_1}
\end{equation}
where \(S_{D}\), \(S_{C}\)  and \(S_{T}\) are the set of URLLC, CL and TS slices respectively. The objective function in \eqref{eq:min_problem_1} is the long-term average power spent by the BS. The first constraint is a reliability condition ensuring that delay is less than $D_i^{\text{max}}$ with a reliability of at least $1 - \gamma_i$, thereby constituting the primary constraint for ensuring URLLC. The rate constraint is addressed implicitly using \eqref{eq:shannon formula} and explicitly in the second and third constraints. And in the last constraint, we assume that each subchannel is allocated to at most one user.  Using this formulation, the instantaneous transmitted power on the $\textit{j-th}$ subchannel is $p_j(t)=\sum_{i=1}^{N}p_{ij}(t)$. 

The problem in \eqref{eq:min_problem_1} is a stochastic mixed-integer linear programming optimization problem, which is NP-hard\cite{b8}. Moreover, as the number of users in the system varies over time, it is challenging to maintain slice isolation following a sudden change in the number of users. Thus, to simplify the problem, we convert the target outage delay constraint in \eqref{eq:min_problem_1} to a target rate constraint $r_{o,i}$. To that end, we find the probability with which the delay of URLLC user $i$ exceeds a threshold $D_i^{max}$, i.e., $\Pr\left(D_{i} > D_i^{max}\right)$. Assuming an M/M/1 queue, it can be proved that 
\begin{equation*}
\begin{aligned}
\Pr(D_i > D_i^{max}) &= e^{-(r_i(t)-a_i(t))D_i^{max}},
& \forall i \in \mathcal{N}_{s},\: \forall s \in \mathcal{S}_{D},
\end{aligned}
\end{equation*}
where $a_i(t) = \zeta_i(t)\xi_i(t)$ is the instantaneous arrival rate in bits per second (bps) for user $i$ which captures the instantaneous packet size $\zeta_i$ and the instantaneous packet inter-arrival rate $\xi_i$ of URLLC user $i$ at time slot $t$. It follows then that
\begin{equation*}
\begin{aligned}
    e^{-(r_i(t)-a_i(t))D_i^{max}} < 1 - \gamma_i .
\end{aligned}
\end{equation*}
Finally, the rate constraint associated to a given outage delay of user $i$ at an instant of time can be written as
\begin{equation}
    r_{o,i}(t)=a_i(t)-\frac{\ln(1-\gamma_i)}{D_i^{max}}.
    \label{eq:target rate_RLL}
\end{equation}
If we consider that the $i$-th user traffic has jitter constraints $J_i$, we can use the expression of the standard deviation of the exponential distribution assumed for $D$
\begin{equation}
    std\{D\}=\frac{1}{r_{o,i}(t)-a_s(t)}.
\end{equation}
Therefore, a joint delay and jitter bound for a URLLC user yields the following required rate 
\begin{equation}
\begin{aligned}
r_{o,i}(t) &= a_i(t) + \max\left(J^{-1}_{i}, \frac{-\ln(1-\gamma_i)}{D_i^{max}}\right), \\
& \quad \forall i \in \mathcal{N}_{s}, \quad \forall s \in \mathcal{S}_{D}, \quad \forall t.
    \label{eq:target rate_RLL2}
\end{aligned}
\end{equation}

Also, knowing the packet size $L$ and the scheduling period $T_{sp}$, the transmission rate of a user $i$ in a TS slice is given by
    \begin{equation}
    \begin{aligned}
    r_{o,i}=L/T_{sp}, \quad \forall i \in \mathcal{N}_{s}, \quad \forall s \in \mathcal{S}_{T}.
    \end{aligned}
    \label{eq:TS_slice_rate}
    \end{equation}

By imposing a rate constraint at each time instant, the average power optimization can be decoupled into instantaneous power optimization. And the target transmission rate for any user $i$ is given by \(r_{o,i}\). We then recast the problem in \eqref{eq:min_problem_1} as
\begin{equation}
\begin{aligned}
\min_{\p,\x}&\sum_{i=1}^{N}\sum_{j=1}^{K}p_{ij},\\&\text{s.t. }r_i\geq r_{o,i},\:\forall i\in \mathcal{N},\\&p_{ij}\geq0,\:x_{ij}\in\{0,1\},\:\forall i\in \mathcal{N},\:\forall j\in \mathcal{K},\\&
\sum_{i=1}^{N}x_{ij}\leq 1, \:\forall j\in\mathcal{K}.
\end{aligned}
\label{eq:min problem 2}
\end{equation}
The minimisation of the instantaneous power in \eqref{eq:min problem 2} will be solved at every time slot or scheduling period. In this way, we may allocate resources in a mixed-traffic scenario.

\section{Radio resource allocation}
\subsection {Power}
Problem \eqref{eq:min problem 2} is convex and feasible if we relax $x_{ij}$ to take values in the interval $[0,1]$, so we rewrite the problem as $\min_{\mathbf{p},\mathbf{x}}\:\mathcal{L}$, where

\begin{equation}
\begin{aligned}
\mathcal{L} &= \sum_{i=1}^{N}\sum_{j=1}^{K}p_{ij} + \sum_{j=1}^{K}\mu_j\left(\sum_{i=1}^{N}x_{ij}-1\right) + \sum_{i=1}^{N}\lambda_i\left(r_{o,i}-r_i\right) \\
&= \sum_{i=1}^{N}\left[\sum_{j=1}^{K}\left(p_{ij} + \mu_jx_{ij}\right) + \lambda_i\left(r_{o,i} - r_i\right)\right] - \sum_{j=1}^{K}\mu_j \\
&= \sum_{i=1}^{N}\mathcal{L}_i(\mathbf{x},\mathbf{p};\boldsymbol{\lambda},\boldsymbol{\mu}) - \sum_{j=1}^{K}\mu_j,
\end{aligned}
\label{eq:lagrangian}
\end{equation}
and $\lambda_i$ and $\mu_j$ are dual variables associated with the rate and subchannel allocation constraints, respectively. By deriving $ \mathcal{L}_i$ with respect to $p_{ij}$ we get the optimum power allocation for the $i$-th user:
\begin{equation}
p_{ij}^*=x_{ij}\left[\frac{\lambda_iB}{\ln2}-\frac{\sigma^2}{h_{ij}}\right]^+.
\label{eq:power allocation}
\end{equation}

\subsection {Subchannels}
To obtain the optimum variables $x_{ij}$ we plug \eqref{eq:power allocation} back into $\mathcal{L}_i(\mathbf{x}, \mathbf{p}; \boldsymbol{\lambda}, \boldsymbol{\mu})$

\begin{equation}
\begin{aligned}
\mathcal{L}_i(\mathbf{x}, & \mathbf{p}^*; \boldsymbol{\lambda}, \boldsymbol{\mu}) = \\
&= \sum_{j=1}^{K}x_{ij}\left(\left[\frac{\lambda_iB}{\ln 2} - \frac{\sigma^2}{h_{ij}}\right]^+ + \mu_j\right) + \lambda_i r_{o,i} \\
&- \lambda_iB \sum_{j=1}^{K}x_{ij} \log_2\left(1 + \left[\frac{\lambda_iB}{\ln 2} - \frac{\sigma^2}{h_{ij}}\right]^+ \frac{h_{ij}}{\sigma^2}\right) \\
&= \lambda_i r_{o,i} - \sum_{j=1}^{K}x_{ij} \left(\mu_{ij}(\lambda_i) - \mu_j\right),
\end{aligned}
\label{eq:reducedpartiallagrangian}
\end{equation}
where we define the auxiliary variable 
\begin{equation*}
\begin{aligned}
\mu_{ij}(\lambda_i) &= \lambda_iB\log_2\left(1+\left[\frac{\lambda_iB}{\ln2}-\frac{\sigma^2}{h_{ij}}\right]^+\frac{h_{ij}}{\sigma^2}\right) \\
&\quad -\left[\frac{\lambda_iB}{\ln2}-\frac{\sigma^2}{h_{ij}}\right]^+
\end{aligned}
\end{equation*}

Therefore, equation \eqref{eq:reducedpartiallagrangian} is minimized with respect to $x_{ij}$ if the following rule is adopted:
\begin{equation}
x_{ij}^*= \left\{\begin{array}{ccl}
1 & \mbox{if}
& \mu_{ij}(\lambda_i)>\mu_j, \\ 0 & \mbox{if} & \mu_{ij}(\lambda_i)<\mu_j, \\
\left[0,1\right] & \mbox{if} & \mu_{ij}(\lambda_i)=\mu_j.
\end{array}\right.
\label{eq:threeconditions}
\end{equation}
Equation \eqref{eq:threeconditions} is the solution to the relaxed problem. Note that we get binary values for the first two cases. As per the third case, we adopt $x_{ij}=1$ as a possible solution, acknowledging some gap to the optimum power sum.

Finally, we get \(\mathcal{L}_i(\boldsymbol{\lambda}, \boldsymbol{\mu}) = \lambda_i r_{o,i} - \sum_{j=1}^{K} \left(\mu_{ij}(\lambda_i) - \mu_j\right)^+\), and from \eqref{eq:lagrangian} the dual function that must be maximized is
\begin{equation}
\mathcal{L}(\boldsymbol{\lambda}, \boldsymbol{\mu}) = \sum_{i=1}^{N} \lambda_i r_{o,i} - \sum_{j=1}^{K} \left[ \sum_{i=1}^{N} \left(\mu_{ij}(\lambda_i) - \mu_j\right)^+ + \mu_j \right].
\label{eq:reducedduallagrangian}
\end{equation}
It can be proved (see appendix I) that \(\mu_j^* = \max_i \mu_{ij}(\lambda_i)\) minimizes the right term in \eqref{eq:reducedduallagrangian}, and hence
\begin{equation}
\mathcal{L}(\boldsymbol{\lambda}) = \sum_{i=1}^{N} \lambda_i r_{o,i} - \sum_{j=1}^{K} \max_i \mu_{ij}(\lambda_i).
\label{eq:primalonlambda}
\end{equation}

Note that the first condition in \eqref{eq:threeconditions} will never occur. Additionally, the third condition states that any value for $x_{ij}$ is valid for the selected user in the \textit{j}-th subchannel. Maximizing this expression requires a multi-dimensional search over $\lambda_1,\dots,\lambda_N$. Since \eqref{eq:primalonlambda} is concave (the dual function is the point-wise infimum of a family of affine functions of $\lambda_i$) in the maximization we can apply subgradient principles in the ellipsoid method \cite{b15}, which has the advantage of the fast convergence and no tuning of a learning rate. The subgradient of \eqref{eq:primalonlambda} reads

\begin{equation}
\begin{aligned}
\frac{\partial \mathcal{L}(\boldsymbol{\lambda})}{\partial \lambda_i} &= r_{o,i} - \sum_{j \in \mathcal{J}_{i}} \frac{\partial \mu_{ij}(\lambda_i)}{\partial \lambda_i}, \\
\frac{\partial \mu_{ij}(\lambda_i)}{\partial \lambda_i} &= 
\begin{cases}
B \log_2\left(\frac{\lambda_i B}{\ln 2} \frac{h_{ij}}{\sigma^2}\right) & \text{if } \lambda_i > \frac{\sigma^2 \ln 2}{B h_{ij}}, \\
0 & \text{else}.
\end{cases}
\end{aligned}
\label{eq:subgradient}
\end{equation}
where \(\mathcal{J}_i = \{j \in \mathcal{K} \mid i^*(j) = i\}\) is the set of subchannels allocated to user \(i\) according to \eqref{eq:threeconditions}. A method to obtain \(\boldsymbol{\lambda}\) is described in algorithm \ref{alg:ellipsoid}.

\begin{algorithm}
\caption{Subgradient Update of $\boldsymbol{\lambda}$ - Ellipsoid Algorithm}
\label{alg:ellipsoid}
\begin{algorithmic}[1]
\State \textbf{Input:} $\boldsymbol{\lambda}, h_{ij}/\sigma^2, r_{o,i}$ for $i=1,\ldots,N$
\State \textbf{Output:} $\boldsymbol{\lambda}, \mathcal{J}_i$ for $i=1,\ldots,N$
\State $n \gets 0, \: (\boldsymbol{D}(0), \boldsymbol{\lambda}(0)) = \text{initial ellipsoid}, \quad \boldsymbol{g}(0) = \text{initial subgradient}$
\While{$\sqrt{\boldsymbol{g}^T  \boldsymbol{D}  \boldsymbol{g}} > \epsilon$}
\For{$j=1:K$}
\State $bestUser(j) = \arg\max_i \mu_{ij}(\boldsymbol{\lambda})$
\EndFor
\For{$i=1:N$}
\State $\mathcal{J}_{i} = \text{listAllocatedRB}(i, bestUser)$
\EndFor
\State $\boldsymbol{g} = \text{gradient} (\boldsymbol{\lambda}, \mathcal{J}_{1}, \dots, \mathcal{J}_{N})$ in \eqref{eq:subgradient}
\State $\boldsymbol{\tilde{g}} = \frac{1}{\sqrt{\boldsymbol{g}^T  \boldsymbol{D}  \boldsymbol{g}}}  \boldsymbol{g}$
\State $\boldsymbol{\lambda} \leftarrow \max(\boldsymbol{\lambda} + \frac{1}{N+1}  \boldsymbol{D}  \boldsymbol{g}, 0)$
\State $\boldsymbol{D} = \frac{N^2}{N^2 - 1} \left(\boldsymbol{D} - \frac{2}{N+1}  \boldsymbol{D}  \boldsymbol{\tilde{g}}  \boldsymbol{\tilde{g}}^T  \boldsymbol{D} \right)$
\State $n \gets n + 1$
\EndWhile
\end{algorithmic}
\end{algorithm}

A proper initialization of algorithm \ref{alg:ellipsoid} requires that the solution for $\lamlam$ is included in the ellipsoid defined by $\mathbf{D}$. If we can upper bound the dual variable as $\lambda_i^*\leq\lambda^u_i$, then a reasonable initial value is $\mathbf{D}=\sum_{i}{\lambda_i^{u}}^2\mathbf{I}$. By recognising that the dual problem is concave and non-linear, equating \eqref{eq:subgradient} to zero is a necessary and sufficient condition to obtain $\lambda_i^*$

\begin{equation}
    \lambda_i^*=2^\frac{r_{oi}}{B|\mathcal{J}_i|}\left(\prod_{j\in\mathcal{J}_i}\frac{\sigma^2}{h_{ij}}\right)^{\frac1{|\mathcal{J}_i|}}\frac{\ln2}{B}
    \label{closed_form_update_lambda}
\end{equation}

Note however that we cannot obtain $\lamlam^*$ in closed form since it requires $\mathcal{J}_i$, but we can upper bound it by applying the inequality between the arithmetic and the geometric means, extend the summation to all subchannels and take $|\mathcal{J}_i|=1$
\begin{equation}
    \lambda_i^*\leq2^\frac{r_{oi}}{B|\mathcal{J}_i|}\frac{\ln2}{B|\mathcal{J}_i|}\sum_{j\in\mathcal{J}_i}\frac{\sigma^2}{h_{ij}}\leq2{^\frac{r_{oi}}{B}}\frac{\ln2}{B}\sum_{j=1}^{K}\frac{\sigma^2}{h_{ij}}=\lambda_i^u
    \label{closed_form_lambda}
\end{equation}
This bound provides an initial $\mathbf{D}$ given a scenario of transmission rates, number of subchannels and channel gains.

\subsection{Guaranteed Slice Isolation}
\begin{figure}[htbp]
\normalsize
\centerline{\includegraphics[width=0.5\textwidth]{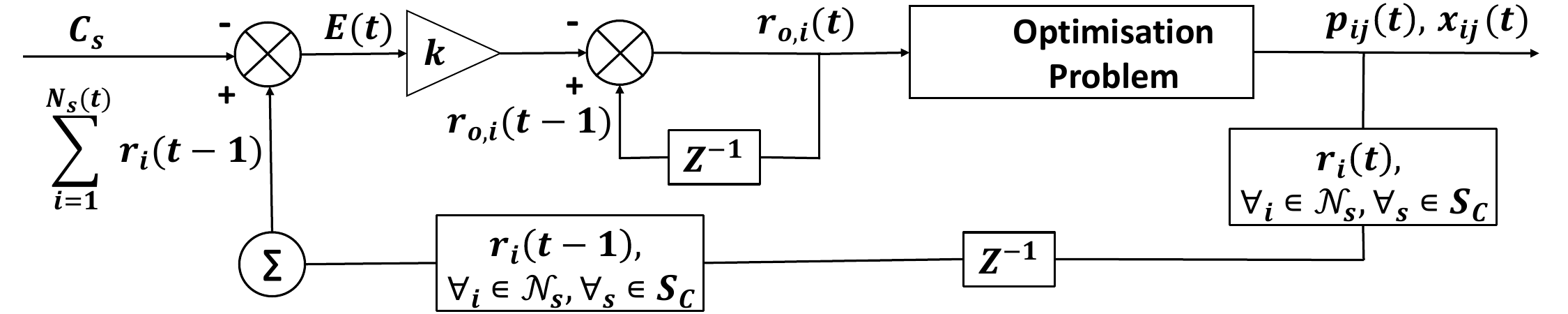}}
\caption{Optimisation control loop for the CL slice}
\label{fig:controller}
\end{figure}

In \eqref{eq:min problem 2}, each user has a rate constraint associated with some \ac{QoS}. If all constraints are met, the resource allocation framework isolates the service of one user from the rest. To this end, we model \eqref{eq:min problem 2} as a control system as shown in Fig.\ref{fig:controller}. The controller is designed to adjust the rate allocation among users in a network slice so that the total transmission rate adheres to a set limit. This approach guarantees slice isolation for CL slices, ensuring that each CL slice operates within its capacity limits without affecting others. The controller generates control signal $\boldsymbol{r_o}$ such that $\boldsymbol{r_o} = [r_{o,1} \; r_{o,2} \; \ldots \; r_{o,Ns}]^T$  is the input parameter vector that controls the solution of the optimisation problem. Where $r_{o,i}(t)$, is the target rate at time slot $t$ for a user $i$ in a CL slice. Thus, we assume that an external agent, based on a control system, recommends an instantaneous value for $r_{o,i}$ according to the observed rate of the delivered traffic.
\\
The control system equation reads
\begin{equation}
r_{o,i}(t) = r_{o,i}(t-1) - k\left(\sum_{i=1}^{N_s(t)} r_i(t-1) - C_s\right).
\label{eq:controller_eqn}
\end{equation}
At each time slot, the controller first calculates the deviation of the actual sum rate from the desired capacity limit, given by $\sum_{i=1}^{N_s(t)} r_i(t-1) - C_s$. This represents the error in the total rate allocation at the current time, $t-1$ for the given CL slice $s$. The controller then adjusts the target rate $\boldsymbol{r_o}(t-1)$ based on this error scaled by $k<1/N_s(t)$, and the direction of the adjustment depends on whether the actual rate is above or below the capacity limit. This adjusted rate $\boldsymbol{r_o}(t)$ becomes the new target rate for the next optimization at time $t$, helping to gradually steer the total rate of the CL slice towards $C_s$.

Equations \eqref{eq:target rate_RLL2} and \eqref{eq:TS_slice_rate} give close form expressions for the target rates of URLLC and TS users respectively. At every time slot, we solve \eqref{eq:target rate_RLL2} and \eqref{eq:TS_slice_rate} for these users which implicitly guarantee slice isolation for URLLC and TS slices. 

\subsection{Admission Control Strategy - Rate Readjustments}
The objective of our optimization problem is to minimize power usage while maintaining the target rate requirements for all users in each network slice. When transmission power requirements exceed the BS available power due to high target rates, it is necessary to readjust the target rates of users in the CL slice to ensure that high-priority tasks in the URLLC and TS slices continue without interruption.

Each user is associated with a Lagrange multiplier, \(\lambda_i\) during optimization as seen algorithm \ref{alg:ellipsoid}. \(\boldsymbol{\lambda}\) indicates how much the overall power consumption will change if the rate constraints are relaxed or tightened. We employ a rate readjustment strategy \cite{b16} detailed in algorithm \ref{alg:QoSreadjustments}, to proportionally lower the target rates across users in the CL slice using their multipliers. The algorithm adopts the well known Newton-Raphson method \cite{b17} to iteratively calculate the difference between available and required power based on \(\boldsymbol{\lambda}\), efficiently reducing rates step by step. The adjustments continue until power consumption is just below the maximum allowable limit, managed by a tolerance value . 

The proportional rate reduction \( \Delta r_{o,i} \) for each CL slice user \( i \) is given by
\begin{equation}
    \Delta r_{o,i} = (p_{\text{opt}} - 1) \left(\frac{\lambda_i}{\sum_{i=1}^N \lambda_i}\right), \ \forall i \in \mathcal{N}_{s}, \ \forall s \in \mathcal{S}_{C}
\end{equation}
where \( p_{\text{opt}} = p_{\text{required}}/p_{\text{available}} \). The rate update formula is thus given by
\begin{equation}
    r_{o,i} = \max(r_{o,i} - \Delta r_{o,i}, 0), \ \forall i \in \mathcal{N}_{s}, \ \forall s \in \mathcal{S}_{C}
    \label{eq:rateupdate}
\end{equation}
By updating the rates with \eqref{eq:rateupdate}, users in CL slice contributing more to the power excess (higher $\lambda_i$) will experience a larger target rate reduction. Thereby efficiently reducing the total power by targeting the reductions more strategically based on users contributions to the power overshoot.

\begin{algorithm}
\caption{Rate Readjustment Algorithm}
\label{alg:QoSreadjustments}
\begin{algorithmic}[1]
\State \textbf{Input:} $r_{o,i}$ $\forall i \in \mathcal{N}_s, \forall s \in \mathcal{S}_C$, and $p_{\text{available}}$
\State \textbf{Output:} $r_{o,i}$ $\forall i \in \mathcal{N}_s, \forall s \in \mathcal{S}_C$, and $p_{ij}$
\State \textbf{Solve} optimization problem (\ref{eq:min problem 2}) and obtain:
\Statex $p_{ij}$, and $\lambda_i$ $\forall i \in \mathcal{N}_s$, $\forall s \in \mathcal{S}_C$
\State \textbf{Evaluate} $p_{\text{required}} = \sum p_{ij}$, and $p_{\text{opt}} = \frac{p_{\text{required}}}{p_{\text{available}}}$
\While{$p_{\text{opt}} > 1 + \epsilon$}
    \State $\Delta r_{o,i} = (p_{\text{opt}} - 1) \left(\frac{\lambda_i}{\sum_{i=1}^N \lambda_i}\right)$
    \State $r_{o,i} = \max(r_{o,i} - \Delta r_{o,i}, 0)$
    \State \textbf{Solve} optimization problem (\ref{eq:min problem 2}) and obtain:
    \Statex \hspace{\algorithmicindent} $p_{ij}$, and $\lambda_i$ $\forall i \in \mathcal{N}_s$, $\forall s \in \mathcal{S}_C$
    \State \textbf{Evaluate} $p_{\text{required}} = \sum p_{ij}$, and $p_{\text{opt}} = \frac{p_{\text{required}}}{p_{\text{available}}}$
\EndWhile
\end{algorithmic}
\end{algorithm}

\section{Simulation Results and Analysis}
\subsection{Scenario Description}
For our simulation, we consider an indoor smart factory environment, characterised by densely packed reflecting elements of diverse shapes and sizes, and different user terminals each with varying QoS requirements that generate different traffic patterns. We consider a 5G BS deployed in this environment with a coverage area radius of 100 m having three slices, one each of  CL, URLLC and TS slices. We use the 3GPP \cite{b18} non-line-of-sight (NLOS) path loss model for indoor factory (InF) dense clutter, low BS (InF-DL) scenario given in \eqref{eq:PL_INF_DL} in combination with Rayleigh fading, to obtain the channel gain in the links between the users and the BS
\begin{equation}
    \begin{aligned}
        PL_{\text{InF-DL}} &= 18.6 + 35.7 \log_{10}d_{3D} + 20 \log_{10}f_c, \\
        PL_{\text{NLOS}} &= \max(PL_\text{InF-DL}, PL_{\text{LOS}}, PL_{\text{InF-SL}}), \\
        \sigma_{\text{SF}} &= 7.2
    \end{aligned}
    \label{eq:PL_INF_DL}
\end{equation}
where \( PL_{\text{LOS}} \) and \( PL_{\text{InF-SL}} \) are respectively the path loss models for InF line-of-sight and NLOS sparse clutter, low BS, in dB. \( f_c \) is the centre frequency in GHz, \( \sigma_{\text{SF}} \) is the shadow fading (SF) standard deviation in dB. The SF is randomly sampled from a lognormal distribution with zero mean. \( d_{3D} \) represents the three-dimensional distance in meters (m) between the transmitter and the receiver and lies in the range \( 1 \leq d_{3D} \leq 100 \) m. Table \ref{tab:Link_Parameters} shows the link parameters assumed for our simulation. Also, the QoS specifications per slice, and the number of active users in each slice is shown in Table \ref{tab:Qos_Specs}. We assume that all users in the URLLC and TS slices have the same QoS specifications. At time slots 33 and 66, we simulate a sudden decrease and increase, respectively, in the number of users in the CL slice. We assume one scheduling period of 10 ms equals one time slot. The total simulation run time is 100 time slots. 
\begin{table}[ht]
    \centering
    \caption{Simulation parameters}
    \begin{tabular}{|l|l|}
    \hline
        \textbf{Parameters} & \textbf{Values} \\ \hline
        Carrier frequency, \(f_c\) & 3.7 GHz  \\ \hline
        Antenna gains  & \(G_t\)  = \(G_u\)  = 0 dBi  \\ \hline
        Transmit power in DL & 23 dBm  \\ \hline
        Noise spectral density & -174 dBm/Hz  \\ \hline
        Total bandwidth, \(B_w\) & 25 MHz\\  \hline
        Subchannel bandwidth, \(B\) & 180 kHz\\  \hline
        Numerology 0 & 15 kHz SCS\\  \hline
        Number of subchannels & 133\\  \hline
        Cell radius & 100 m\\  \hline
    \end{tabular}
    \label{tab:Link_Parameters}
\end{table}

\begin{table}[ht]
\centering
\caption{Per slice QoS specifications and number of active users}
\label{tab:Qos_Specs}
\renewcommand{\arraystretch}{1.2} % Adjust cell height
\resizebox{\columnwidth}{!}{
\begin{tabular}{|c|>{\raggedright\arraybackslash}p{1.75cm}|>{\centering\arraybackslash}m{1.3cm}|>{\centering\arraybackslash}m{0.85cm}|>{\centering\arraybackslash}m{1.6cm}|>{\centering\arraybackslash}m{0.85cm}|}
\hline
\textbf{Slice} & \textbf{Requirements} & \textbf{Target Rates} & \multicolumn{3}{c|}{\textbf{No. of Active Users at Time Slot t}} \\ \cline{4-6}
 &  &  & \( t < 33 \) & \( 33 \leq t \leq 66 \) & \( t > 66 \) \\ \hline
CL & \( C_s = 27 \, \text{Mbps}\) & 27 Mbps & 5 & 2 & 7 \\ \hline
URLLC & \( J_s = 1 \, \text{ms} \newline \gamma_s = 99.9\% \newline a_s = 2 \, \text{Mbps} \newline D_{s}^{\text{max}} = 10 \, \text{ms} \) & 2 Mbps & 2 & 2 & 2 \\ \hline
TS & \( L = 5 \, \text{kB} \newline T_{sp} = 10 \, \text{ms} \) & 1.64 Mbps & 1 & 1 & 1 \\ \hline
\end{tabular}
}
\end{table}

\subsection{Slice Isolation and QoS Assessments}
\begin{figure}[ht]
    \centering
    \begin{minipage}{0.49\linewidth}
        \centering
        \includegraphics[width=\linewidth]{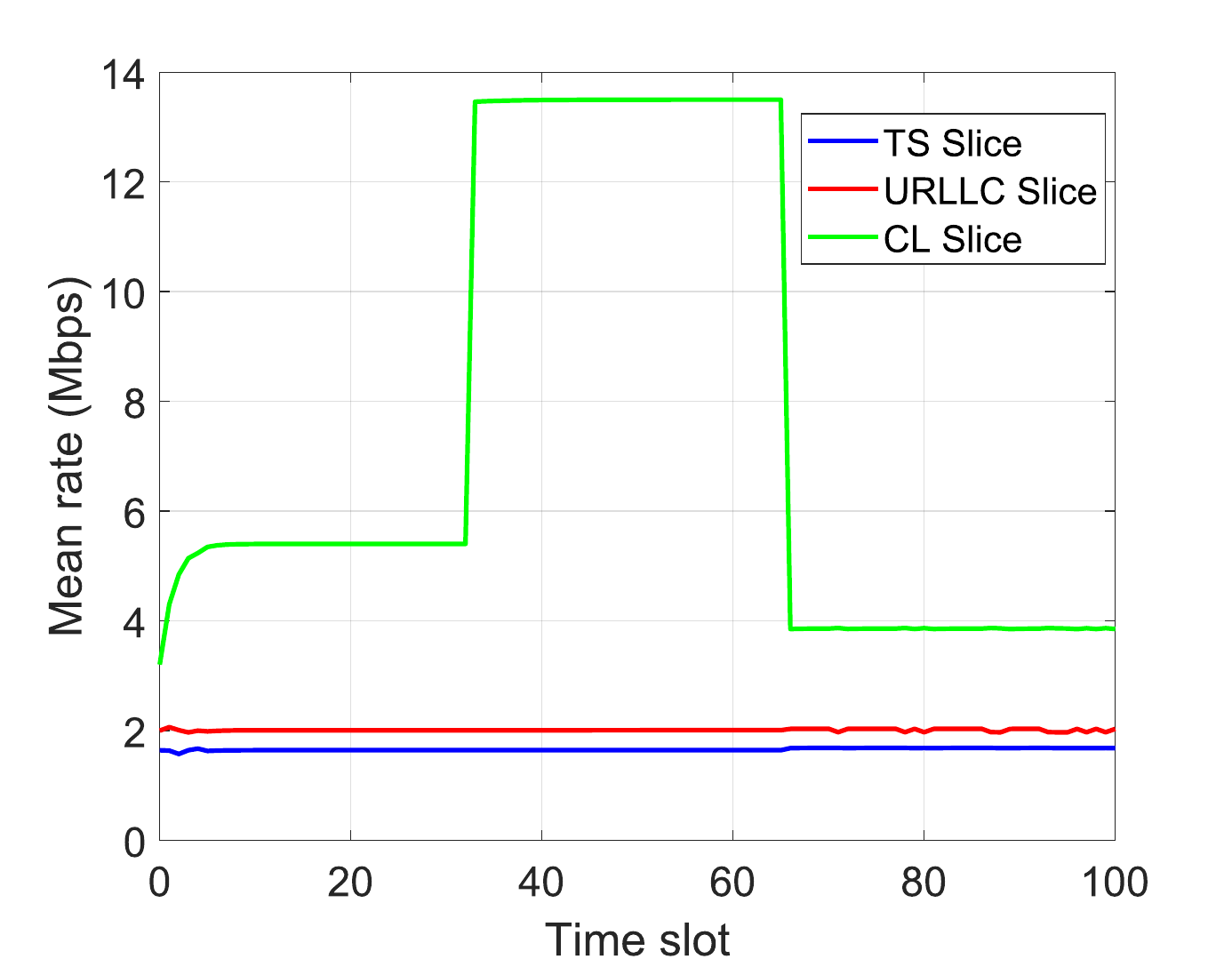}
        \footnotesize{(a) Mean rate of users per slice}
    \end{minipage}
    \hfill
    \begin{minipage}{0.49\linewidth}
        \centering
        \includegraphics[width=\linewidth]{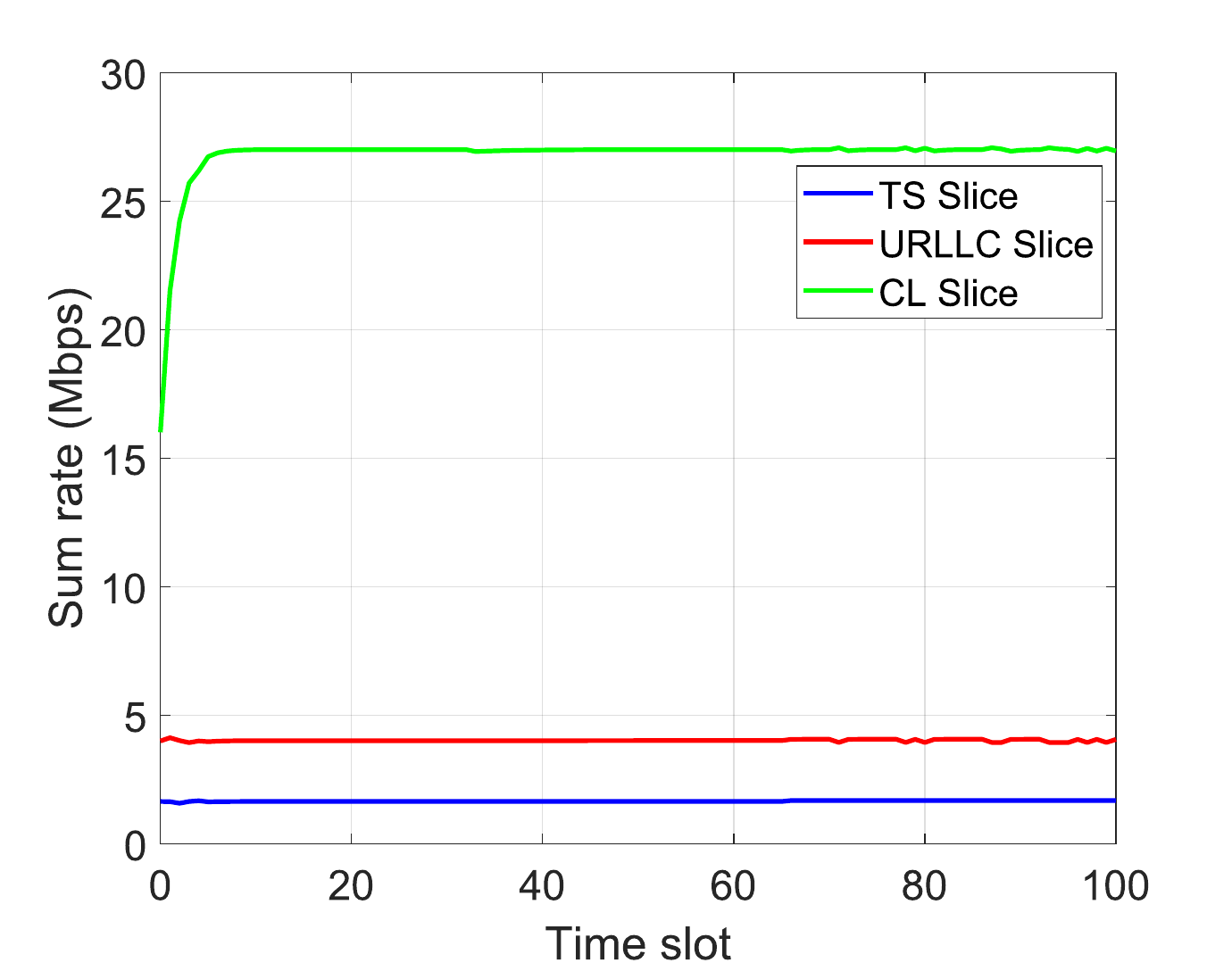}
        \footnotesize{(b) Sum rate of users per slice}
    \end{minipage}
    \caption{Achievement of (a) slice isolation and (b) target QoS for the three slices considered}
    \label{fig:rate_metrics_without_AC}
\end{figure}

To evaluate the achievement of effective slice isolation, we show the mean user rates across the three slices in Fig.\ref{fig:rate_metrics_without_AC}(a). From this figure, we can observe that the mean rate of users in the CL slice experiences abrupt increases and decreases in response to the sudden changes in the number of users, decreasing and increasing respectively, at those specific times. However, the mean rate for the users in the URLLC and TS slices remains consistent, indicating complete isolation between the slices.

 Given that all QoS constraints where translated into rate constraints, meeting the target rates for each slice equates to fulfilling their respective QoS requirements. Fig.\ref{fig:rate_metrics_without_AC}(b)  shows the sum rates of users in each slice. This figure clearly indicates that all the target rates per slice, as outlined in Table \ref{tab:Qos_Specs}, have been met, thereby demonstrating the achievement of the delay, jitter, reliability and bitrate constraints for all users. An initial transient for the sum rate in the CL slice is observed as a result of the transient behaviour in the control loop in \eqref{eq:controller_eqn}.

%%%%%%%%%%%%%%%%%%%%%%%%%%%%%%%%%
%%%%%%%%%%%%%%%%%%%%%%%%%%%%%%%%%%%%%%%%%%

\subsection{Admission Control}
To evaluate the effectiveness of our admission control strategy, we simulate periods of network congestion by increasing both the number of users and their target rates in the slices, as detailed in Table \ref{tab:Qos_Specs2}.

\begin{table}[ht]
\centering
\caption{Per Slice Target Rates and Number of Active Users during Network Congestion}
\label{tab:Qos_Specs2}
\renewcommand{\arraystretch}{1.2} % Adjust cell height
\scriptsize % Reduce font size
\resizebox{\columnwidth}{!}{ % Resize table to fit within column boundaries
\begin{tabular}{|c|>{\centering\arraybackslash}m{1.5cm}|>{\centering\arraybackslash}m{0.8cm}|>{\centering\arraybackslash}m{1.6cm}|>{\centering\arraybackslash}m{0.8cm}|}
\hline
\textbf{Slice} & \textbf{New Target Rates} & \multicolumn{3}{c|}{\textbf{No. of Active Users at Time Slot t}} \\ \cline{3-5}
               &                           & \(t < 33\) & \(33 \leq t \leq 66\) & \(t > 66\) \\ \hline
CL             & 270 Mbps                 & 10                  & 7                             & 12                  \\ \hline
URLLC          & 20 Mbps                  & 4                   & 4                             & 4                   \\ \hline
TS             & 1.64 Mbps                & 2                   & 2                             & 2                   \\ \hline
\end{tabular}
}
\end{table}

\begin{figure}[htbp]
    \centering
    \begin{minipage}{0.49\linewidth}
        \centering
        \includegraphics[width=\linewidth]{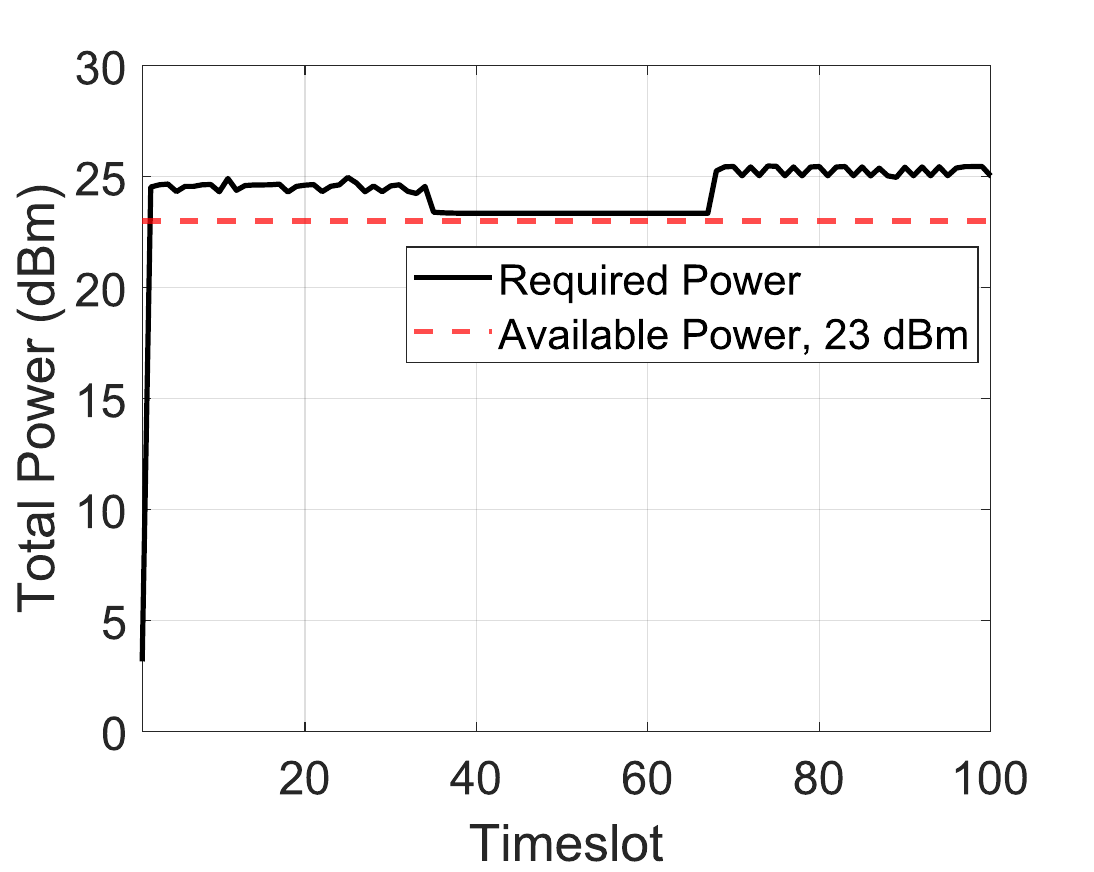}
        \footnotesize{(a)}
    \end{minipage}
    \hfill
    \begin{minipage}{0.49\linewidth}
        \centering
        \includegraphics[width=\linewidth]{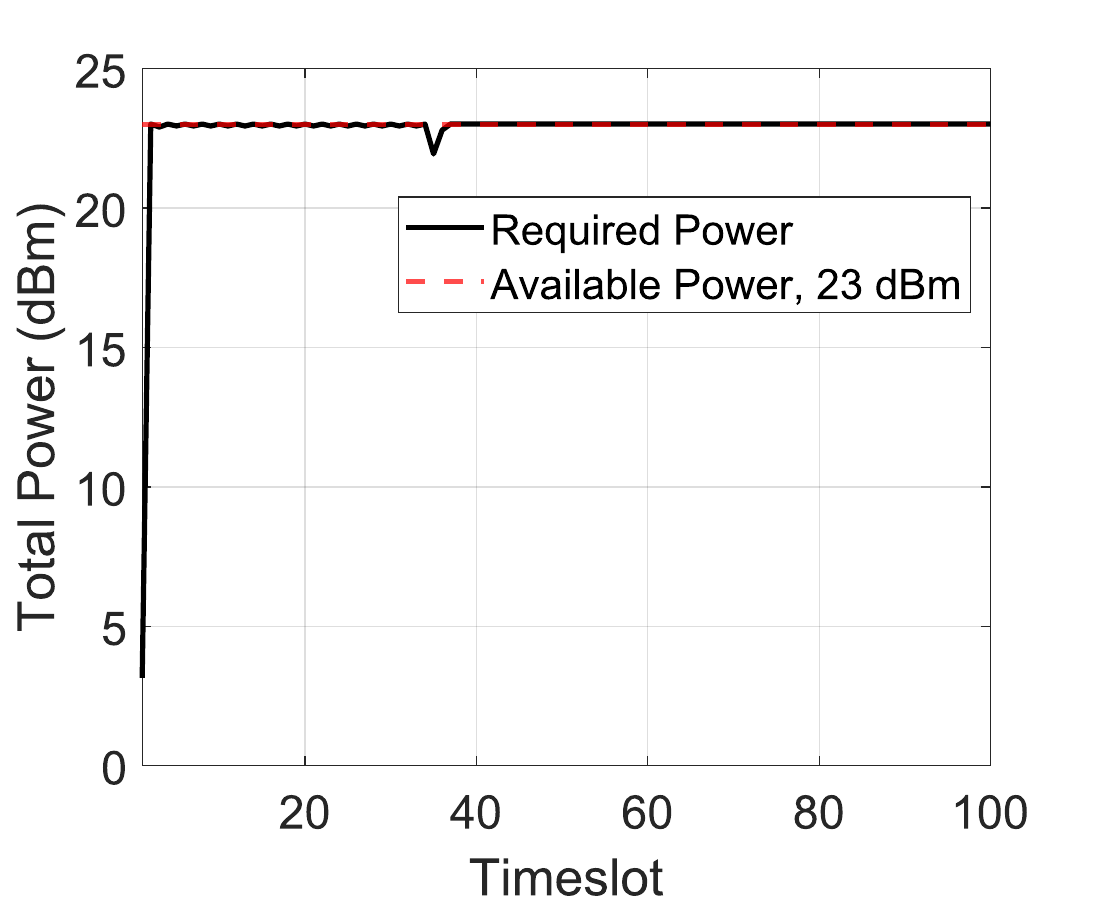}
        \footnotesize{(b)}
    \end{minipage}
    \caption{BS transmit power (a) without and (b) with admission control}
    \label{fig:power_comparison}
\end{figure}

\begin{figure}[htbp]
    \centering
    \begin{minipage}{0.48\linewidth}
        \includegraphics[width=\linewidth]{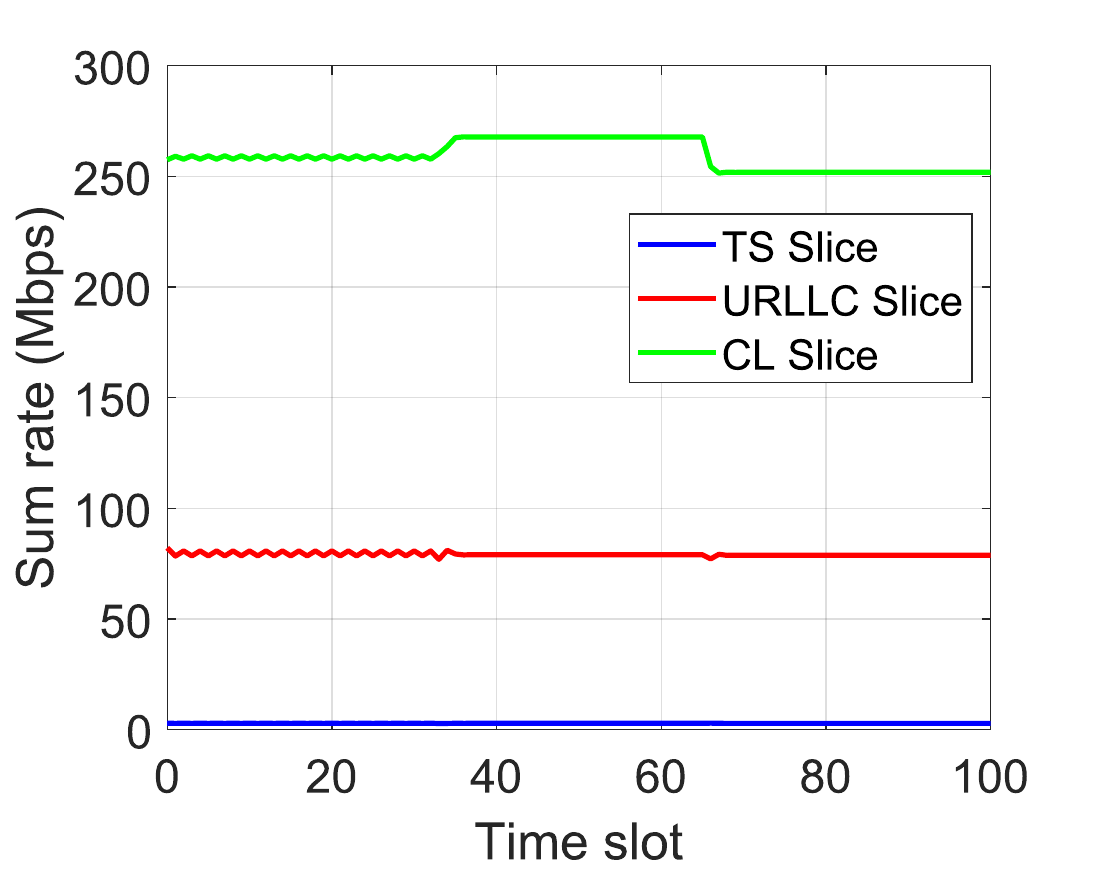}
        \centering
        \footnotesize{(a)}
    \end{minipage}
    \hfill
    \begin{minipage}{0.48\linewidth}
        \includegraphics[width=\linewidth]{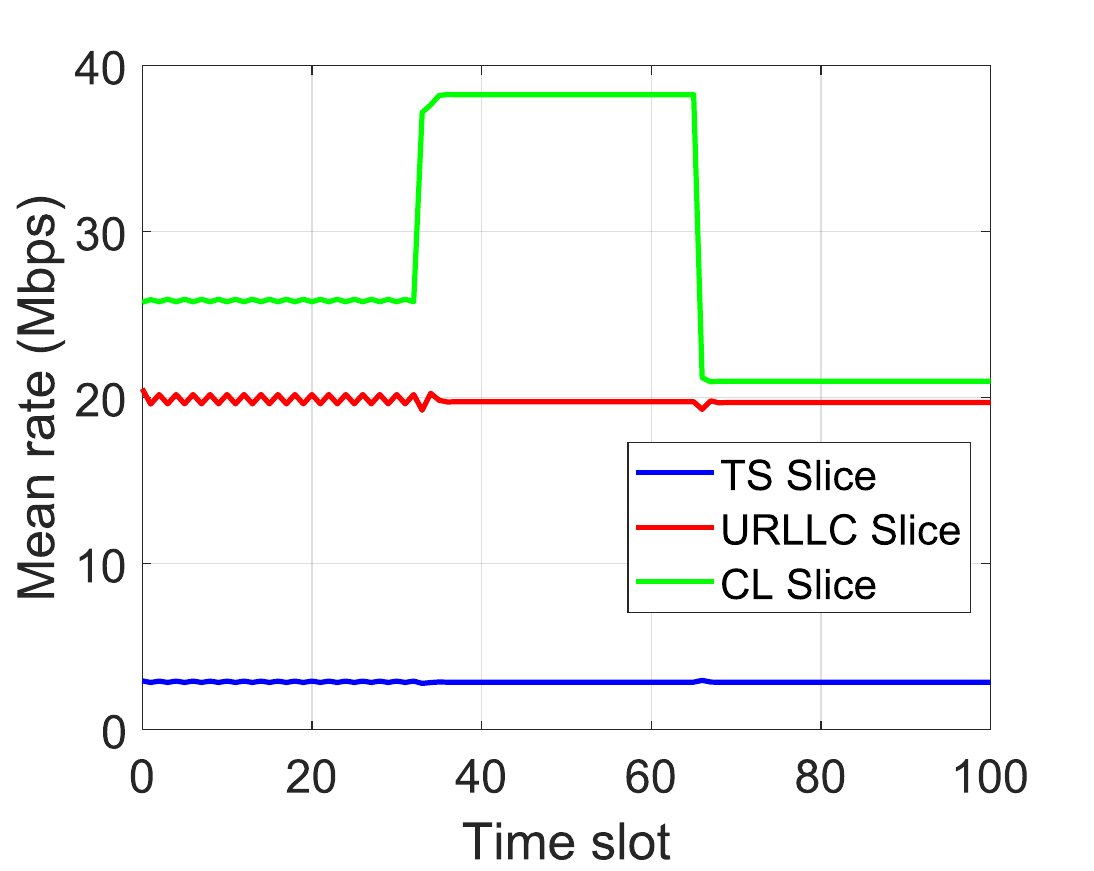}
        \centering
        \footnotesize{(b)}
    \end{minipage}
    \caption{Sum rates (a) and mean rates (b) of users per slice during admission control}
    \label{fig:rates_with_ac}
\end{figure}

Fig.\ref{fig:power_comparison}(a) shows the total BS power without admission control. We can observe that the power required to satisfy the QoS for all users across the three slices exceeds the available 23 dBm even when the number of users in the CL slice decreased at $33 \leq t \leq 66$. This insufficiency indicates that the requirements of all active users will not be met, which can disrupt critical transmissions. To prevent this, we implement admission control to limit the required power to 23 dBm by selectively reducing the rates of users in the CL slice with less stringent requirements. The effects of this strategy are depicted in Fig.\ref{fig:power_comparison}(b) and Fig.\ref{fig:rates_with_ac}.

The total BS power when admission control is active is shown in Fig.\ref{fig:power_comparison}(b). The power requirement remains within the available limit of 23 dBm, confirming the success of our strategy. The quick drop in the total power observed at $t=33$ corresponds to the sudden decrease in the number of users at that time slot. Additionally, Fig.\ref{fig:rates_with_ac}(a) illustrates the rate adjustments in the CL slice during admission control. Specifically, at $t<33$ when the required power initially exceeds 23 dBm, the sum rate of the CL slice is reduced from the target of 270 Mbps to approximately 259 Mbps. At $33\leq t \leq 66$, as the number of users in the CL slice decreases from 10 to 7, the sum rate is readjusted to around 267 Mbps. Furthermore, when the number of users increased from 7 to 12 at $t>66$, the sum rate reduced again to 251 Mbps. 

During these rate readjustments, we can see from Fig.\ref{fig:rates_with_ac}(a) that the sum rates for both the URLLC and TS slices remains unaffected, indicating that their strict QoS requirements are preserved even in periods of network congestion. Correspondingly, the mean rates for users in the URLLC and TS slices continue unchanged during admission control as depicted in Fig.\ref{fig:rates_with_ac}(b). These consistent rates further affirm that our framework guarantees complete slice isolation.

%%%%%%%%%%%%%%%%%%%%%%%%%%%%%%%%%%%%%%%%%

%%%%%%%%%%%%%%%%%%%%%%%%%%%%%%%%%%%%%%%

\section{Conclusion and Future work}
In this paper, we studied a network slice-aware power optimization and subchannel allocation problem in smart factory wireless networks, accommodating varying number of users over time. We considered three types of slices: TS slices handling time-sensitive traffic, URLLC slices with strict latency, jitter, and reliability requirements, and CL slices with capacity demands. We proposed a novel approach for guaranteeing slice isolation and significantly reducing the complexity of the problem formulation which is often encountered in optimisation-based resource allocation techniques. We have shown that our framework minimises power while satisfying the QoS requirments of users in the three slices. And in the event of unaffordable power consumption at the BS, we demonstrate effective admission control to preserve critical connections. Simulation results validate that our approach meets the QoS demands of all slices and maintains slice isolation under changing network conditions.

This research lays the groundwork for developing a dynamic, efficient, and autonomous slice-aware radio resource management system for smart factories. Future work will delve into the slicing resource allocation challenges posed by mobile terminals like AGV and evaluate performance using realistic industrial scenarios with actual factory channel traces and its application to uplink communications.

\section{Appendix I}
Take a sequence $\mu_1\leq\mu_2\leq\dots\leq\mu_{N}$ of real values, and define $F(\mu)$ a function to be minimized with respect to $\mu$
\begin{equation}
\begin{aligned}
F(\mu)&=\sum_{i=1}^{N}\left(\mu_{i}-\mu\right)^++\mu=\sum_{i=i^*}^{N}\left(\mu_{i}-\mu\right)+\mu\\&=\sum_{i=i^*}^{N}\mu_{i}-(N-i^*+1)\mu+\mu=\sum_{i=i^*}^{N}\mu_{i}-(N-i^*)\mu
\end{aligned}
\end{equation}
where \(i^*\) is such that \(\mu_{i^*-1} \leq \mu < \mu_{i^*}\). If \(\mu > \mu_N\), then \(F(\mu) = \mu > \mu_N\). If \(i^* = N\), then \(F(\mu) = \mu_N\). If \(i^* = N-1\), then \(F(\mu) = \mu_{N-1} + \mu_N - \mu\), with \(\mu \in \left[\mu_{N-2}, \mu_{N-1}\right]\), that is \(\mu = \alpha \mu_{N-2} + (1-\alpha) \mu_{N-1}\) with \(\alpha \in \left[0,1\right]\). Hence, \(F(\mu) = \mu_N + \alpha(\mu_{N-1} - \mu_{N-2}) > \mu_N\). The same conclusion can be derived for \(i^* = N-2, N-3, \ldots\). Therefore, the minimum value for \(F(\mu)\) is obtained at \(\mu^* = \mu_N\).

\vspace{12pt}

\end{document}